\documentclass[12pt]{article}
\usepackage{amsmath}
\usepackage{amssymb}

\usepackage{chicago}

\begin{document}


\subsection*{Discussion on Fearnhead P. and D. Prangle (2012). Constructing summary statistics for approximate Bayesian computation: Semi-automatic approximate Bayesian computation, {\it J. Roy. Statist. Soc. B}, {\bf 74} (3), 1-28.}

Dr. D. J. Nott, National University of Singapore, Singapore\\
Dr. Y. Fan, University of New South Wales, Australia\\
Dr. S. A. Sisson\footnote{Communicating Author: {\tt Scott.Sisson@unsw.edu.au}}, University of New South Wales, Australia\\


\noindent 
This paper introduces several novel ideas, including a method of deriving 
summary statistics, $s$,  for functions of model parameters in a way that minimises the variability of the posterior mean of those functions.  
In this comment, we make a connection between this approach to summary statistic choice and Bayes Linear Analysis
(Goldstein and Wooff, 2007).
Bayes Linear Analysis can be viewed as  optimal linear estimation 
of a parameter vector $\theta$ 
where an estimator of the form $a+Bs$ is constructed 
for a $p$-dimensional vector, $a$, and a $p\times d$ matrix, $B$, minimising
\begin{equation}
  E[(\theta-a-Bs)^{\top}(\theta-a-Bs)],  \label{blcriterion}
\end{equation}
where $s$ is a $d$-vector of data (i.e. summary statistics).
The expectation is with respect to the joint prior distribution of $s$ and $\theta$.
The optimal linear estimator is given by
\begin{eqnarray*}
 E_s(\theta)  =  E(\theta)+\mbox{Cov}({\theta,s})\mbox{Var}(s)^{-1}[s-E(s)]. 
\end{eqnarray*}
The estimator 
$E_s(\theta)$ is referred to as the adjusted
expectation of $\theta$ given $s$.  A Monte Carlo approximation to 
(\ref{blcriterion}) based on $(\theta^{(m)},s^{(m)})\sim p(s|\theta)p(\theta)$, $i=1,\ldots,M$, is a least squares criterion
for a linear regression of the simulated parameters on the summary statistics.

Thus, for large $M$, the semi-automatic summary statistics of Fearnhead and Prangle (2012) 
can be viewed as Bayes linear estimates of the posterior means. 
When computations are performed on a restricted parameter space (as per Section 3, point (a)), 
this interpretation still holds under a truncated prior for $\theta$. Finally, the Bayes 
linear interpretation also holds for more flexible regression models, by considering 
suitable basis expansions involving functions of $s$, assuming that transformations 
of $\theta$ to maintain homoscedasticity are 
available. The links between regression methods in ABC and Bayes Linear Analysis 
are discussed further in Nott et al (2011).

Our final comment relates to the identification of a single summary statistic per posterior parameter of interest. In Nott et al. (2011), we propose to improve the accuracy of the joint posterior sample from any ABC method by firstly independently estimating the marginal posteriors $p(\theta_i|s_{obs})$, $i=1,\ldots,p$. Estimating marginal posteriors is easier than estimating the joint posterior due to the lower dimensionality. We then replace the margins of the joint posterior with the more precisely estimated marginal distributions, thereby providing a more precise estimate of the true posterior distribution. This marginal-adjustment strategy will be very efficient if highly informative, but low-dimensional and identifiable summary quantities are available for each marginal parameter.

As such, we propose that our marginal adjustment-strategy using the semi-automatic summary statistics of Fearnhead and Prangle (2012), following a standard ABC analysis using the same statistics, would potentially provide even more precise estimates of the true posterior distribution. This approach is less affected by the increase in dimensionality of $\theta$, than for regular ABC analyses.

\subsection*{Discussion on Fearnhead P. and D. Prangle (2012). Constructing summary statistics for approximate Bayesian computation: Semi-automatic approximate Bayesian computation, {\it J. Roy. Statist. Soc. B}, {\bf 74} (3), 1-28.}

Dr. S. A. Sisson\footnote{Communicating Author: {\tt Scott.Sisson@unsw.edu.au}}, University of New South Wales, Australia\\
Dr. Y. Fan, University of New South Wales, Australia\\


\noindent
This paper proposes a way of deriving summary statistics for functions of model parameters in a way that minimises the variability of the posterior mean of those functions. Based on samples of $(\theta,s)$ in a truncated region of the prior (as per Section 3, point (a)), one fits a regression model e.g. $\theta = \alpha + \beta F(s)$, where $F(s)=(f(s),\ldots,f(s))$. The proposed summary statistic is the regression mean response, $\beta F(s)$, with precisely one statistic for each function of interest.
While using $\beta F(s)$ rather than $s$ allows more precise estimation of the marginal posterior means of the functions of interest, it seems credible that posterior expectations of certain other quantities may be estimated less precisely under $\beta F(s)$ than $s$. This implies that {\it all} posterior expectations of interest in any analysis must be handled in this manner in order to guarantee the best possible precision.

However, in some ABC applications, such as extreme value theory (e.g. Bortot et al, 2007; Erhadt and Smith, 2012), interest is typically in a large number (or even all) posterior quantiles (point estimates and credible intervals) above some high threshold. Our question is how does one mechanistically handle a very large (or even infinite) number of posterior functions of interest within the proposed framework?

In principle, one could use the proposed process directly, and regress all $p'>>p$ posterior quantities of interest against $f(s)$, and using the resulting  $\beta F(s)$ as the relevant summary statistics.  However, as $p'$ becomes large (or even as $p'\rightarrow \infty$, for example, where interest is in all posterior quantiles), this means that the accuracy of the resulting ABC posterior approximation will fall dramatically, compared to when using just $s$, given the increased dimension of the vector of summary statistics $\beta F(s)$. This comes in addition to the required increase in the number of $(\theta, s)$ samples required to perform the regression.
Alternatively, one could repeatedly perform many separate implementations of the proposed procedure, each one aiming to estimate different (lower dimensional) aspects of the posterior as precisely as possible. Of course this approach raises questions of computational overheads, whether the separately estimated quantities would be consistent with eachother, and which combinations of functions of interest to include in each analysis e.g. all posterior parameter means and one function of interest, or some other combination.

Our final comment notes that the performance of regression-based ABC procedures, such as Beaumont et al. (2002), is sensitive to multicolinearity and large numbers of uninformative summary statistics, and as such may ``over-adjust" the $(\theta,s)$ sample and thereby poorly estimate the posterior mean. As the dimension of $f(s)$ would increase rapidly with $p'$, this naturally raises the question as to how the proposed semi-automatic framework would perform in the case of large $p'$ with a potentially unreliable regression component.

\newpage
\subsection*{References}

Beaumont M. A., W. Zhang and D. J. Balding (2002). Approximate Bayesian computation in population genetics. {\it Genetics}, {\bf 162}, 2025-2035.
\\

\noindent Bortot P., S. G. Coles and S. A. Sisson (2007). Inference for stereological extremes. {\it Journal of the American Statistical Association}, {\bf 102}, 84-92.
\\

\noindent Erhardt R. J. and R. L. Smith (2012). Approximate Bayesian computation for spatial extremes. {Computational Statistics and Data Analysis}, in press. {\tt http://arxiv.org/abs/1109.4166}
\\

\noindent
Goldstein, M. and D. Wooff (2007). Bayes Linear Statistics: Theory and Methods. Wiley.\\

\noindent Nott D. J., Y. Fan, L. Marshall and S. A. Sisson (2011). Approximate Bayesian computation and Bayes linear analysis: Towards high-dimensional approximate Bayesian computation. Submitted for publication.  {\tt http://arxiv.org/abs/1112.4755}

\end{document}